\documentclass[aps,prd,epsf,showpacs,amsmath,amssymb,10pt,twocolumn]{revtex4}
\usepackage{d column} 
\usepackage{bm}
\usepackage{color,graphicx}

 \newcommand{\be}{\begin{equation}}
\newcommand{\ee}{\end{equation}}
\newcommand{\ba}{\begin{eqnarray}}
\newcommand{\ea}{\end{eqnarray}}
\newcommand{\ban}{\begin{eqnarray*}}
\newcommand{\ean}{\end{eqnarray*}}

\newcommand{\non}{\nonumber}

\newcommand{\ph}{\ensuremath{{\phi}}}

\begin{document}
\title{Collapse and dispersal in massless scalar field models}
\author{Swastik Bhattacharya$^1$, Rituparno Goswami$^2$, Pankaj S. Joshi$^1$}
\email{swastik@tifr.res.in, Rituparno.Goswami@uct.ac.za, psj@tifr.res.in}
\affiliation{$^1$Tata Institute for Fundamental Research, Homi Bhabha Road,
Mumbai 400005, India}
\affiliation{$^2$Department of Mathematics and Applied Mathematics and Centre for 
Astrophysics, Cosmology and Gravitation,
University of Cape Town, 7701 Rondebosch, Cape Town, South Africa}

\begin{abstract}
The phenomena of collapse and dispersal for a massless scalar field has drawn
considerable interest in recent years, mainly from a numerical perspective. 
We give here a sufficient condition for the dispersal to take place 
for a scalar field that initially begins with a collapse. 
It is shown that the change of the gradient of the scalar field from a timelike 
to a spacelike vector must be necessarily accompanied by the dispersal of the 
scalar field. This result holds independently of any symmetries of the 
spacetime. We demonstrate the result explicitly by means of an example, 
which is the scalar field solution given by Roberts. The implications of 
the result are discussed.

\end{abstract}
\maketitle

\section{Introduction}

The spherically symmetric collapse of a massless scalar field has been of much 
interest towards understanding the dynamical evolutions in general relativity. 
Both analytical 
\cite{Christ} 
and numerical 
\cite{Choptuik} 
investigations have been undertaken by various authors to gain more insight 
into the formation of black holes. One remarkable finding of these 
numerical investigations 
is the demonstration of criticality in gravitational collapse. 
Specifically, it was found that for a range of  
values of the parameter characterizing the solution, black hole forms and there 
was a critical value of the parameter beyond which the solutions are such that 
the scalar field disperses without forming any black hole. 
However, this result has been 
obtained mainly through numerical studies and a proper theoretical understanding 
of this phenomenon is still lacking 
(see e.g. \cite{Gundlach} and the references therein).

Therefore, it is important to understand better the phenomena 
of collapse 
and dispersal analytically. Since very few solutions 
of the massless scalar 
field coupled to gravity are known, it is not possible to reach any conclusion 
in this case only by
studying such particular solutions. 
To obtain more insight into this, it is necessary then 
to identify some general features associated with the collapse 
and dispersal of the scalar fields. 
We study here the collapse of the scalar field in the sense that the expansion 
is negative to begin with along a comoving world line, 
and later it turns positive, thus 
causing the dispersal of the field. If this happens 
along all the world lines, the field
enters a dispersal phase as a whole, and in the case 
otherwise the gravitational collapse continues.

Specifically, we give here a sufficient condition 
for the dispersal of scalar field along its 
world line. We show that if along the worldline, the gradient 
of the scalar field approaches a null value from a timelike  
vector, then a dispersal of the scalar field must take 
place somewhere along this world line,
before the gradient changes. 
This result applies to any massless scalar field spacetime in general, and is not 
dependent on any symmetry of the spacetime like spherical 
symmetry or self-similarity etc. 
Here we discuss the dispersal in a local sense to begin 
with (along one world line). 
However, if this condition of dispersal is satisfied 
along all the comoving 
world lines, then the scalar field will globally 
disperse away without forming a black hole.
We demonstrate explicitly that this is actually the case in the 
solution given by Roberts \cite{Roberts}, Brady \cite{Brady} and Oshiro et al 
\cite{Oshiro}. We analyze this solution here in some detail along these 
lines, which provides some interesting new information on its 
structure.
Therefore, in some classes of models, this condition can 
be used to get more insight into the phenomena of global dispersal 
of a scalar field. 
The above authors found qualitatively different behaviour 
of the solution for different ranges of values of a certain parameter $p$, which 
characterizes the solution. Specifically for $p>1$, black hole forms due to 
collapse, whereas for $0<p<1$ the scalar field first collapses 
and then disperses away without forming any black hole. The case 
$p=1$ is the critical case and a null singularity forms 
in this situation. Near the critical point, the black hole 
mass was found to satisfy a power law behaviour.

Here we show, in order to demonstrate our result mentioned above, that 
the gradient of the scalar field necessarily changes from timelike to spacelike 
along any world line in the case $0<p<1$, and then it is 
shown that a dispersal of 
the scalar field occurs in this case. In the case when 
the field goes to a black hole,
no such gradient change takes place. Another interesting 
point to note is that, in this case, the change of gradient is related 
with the global dispersal of the scalar field, in the sense that 
the field as a whole disperses as mentioned above.


We shall use comoving coordinates here. It is useful to note 
the limitation of comoving coordinates for massless 
scalar field spacetimes, and its domain of validity.
We know that in terms of the eigenvectors of the energy-momentum tensor, 
matter fields can be classified in four distinct types 
\cite{ellis}.
However, all known physical 
matter fields in the universe fall under the first two 
(namely {\it Type I} and {\it Type II}). 
Matter fields with non-zero rest mass are of the first type, 
whereas zero rest mass 
fields can either be the first or the second type, depending on the Lagrangian 
of the field.  For any {\it Type I} field we can use comoving coordinates in 
which the energy momentum tensor of the field is diagonal, whereas this is 
not possible for a  {\it Type II} field. We note that a massless scalar field 
can be of either type depending on the gradient of the field. 
If the gradient is timelike,
the field is of the first type and has the equation of 
state of a stiff fluid and if at 
some spacetime point it becomes null, then the scalar field 
becomes a {\it Type II} field. Hence 
a comoving coordinate system would break down, if the gradient 
of the scalar field becomes null. This 
is the reason for the preference of single or double 
null coordinate systems by some 
authors to study dynamical solutions for the massless scalar field.
For our purpose, however, the comoving system is fully adequate as will
be pointed out below.

In the following, we clarify what is meant by dispersal of the scalar field, 
by relating this with the volume expansion coefficient of the scalar field.
We then show in general that the change of the scalar 
field gradient from a timelike to a spacelike vector must be accompanied by the 
dispersal of the field. 
Then we discuss the Roberts solution as an example and 
show how a change of gradient takes place in this model, and point out 
its connection with the dispersal. 
We can use the comoving coordinates in the spacetime patch where the 
gradient of the field is timelike, and show that for 
$0<p<1$, the volume expansion 
for the scalar field indeed changes sign, signalling a bounce.

\section{Collapse condition in comoving frame}

For matter fields which have a timelike four-velocity vector, we can always 
have a 
comoving reference frame where we can define a timelike 
matter congruence locally as,
\begin{equation}
 u^a= \frac{dx^a}{d\tau}\;;\; u^au_a=-1.
\end{equation}
Here $x^a$ denotes the position for the matter and $\tau$ is the proper 
time along the congruence. Given this congruence $u^a$, we can define a unique 
projection tensor as $h^a_b=g^a_b+u^au_b$. This determines the orthogonal metric 
properties of the instantaneous rest spaces of observers 
moving with four-velocity 
$u^a$ and projects all the geometrical quantities to the 
three-space orthogonal to $u^a$
~\cite{ellis}.

The matter energy-momentum tensor can be decomposed 
relative to $u^a$ in the form
\begin{equation}
 T^{ab}= \rho u^a u^b+ q^a u^b+ q^b u^a+ p h^{ab}+ \Pi^{ab}
\end{equation}
where $\rho=T_{ab}u^au^b$ is the {relativistic energy density} relative to 
$u^a$, $q^a=-T_{bc}u^bh^{ca}$ denotes the {relativistic energy flux}, 
$p=\frac13T_{ab}h^{ab}$ is the 
{isotropic pressure}, and $\Pi_{ab}=T_{cd}h^c_{<a}h^d_{b>}$ is the {trace-free 
anisotropic pressure}. The physics of the situation 
is defined in the {\it equation of 
state} relating these thermodynamical quantities of the 
matter field. For an isentropic 
perfect fluid with a general equation of state $p=p(\rho)$, 
we impose the conditions 
$q^a=\Pi_{ab}=0$. The volume expansion for this matter congruence 
is defined as $\theta= h^a_b \nabla_au^b$,
where $\nabla$ denotes the full covariant derivative. 
Also the `{ time {\em (dot)} 
derivative}' of any function $f(x^a)$ is defined as the 
derivative along $u^a$, that is,
$\dot{f}=u^a\nabla_af$. In terms of the above quantities 
the energy conservation law for 
an isentropic perfect fluid is then given by
\begin{equation}
 \dot{\rho}= - \theta (\rho+p). \label{conserv}
\end{equation}

The continual collapse condition in a comoving frame 
can now be defined in the following way:
{\em For the matter field which is undergoing continual 
collapse because of its self 
gravity, the volume element defined by the matter congruence always decreases 
with the proper time along the congruence, that is, the volume expansion 
is negative.} 

The spacetime singularity corresponds to the 
points where this volume element goes to zero. 
This implies that for a continually collapsing 
matter congruence we must have $\theta <0$ along the world lines. 
In that case, for any collapsing 
isentropic perfect fluid obeying the weak energy condition, since we have  
$\rho \ge0$ and $\rho+ p\ge0$, therefore 
we have $\dot{\rho}\ge0$, i.e. the density 
is a non-decreasing function of the proper time along the matter congruence.

\section{A sufficient condition for dispersal}

The energy-momentum tensor of a massless scalar field $\phi(x^a)$, is given by
\begin{equation}
 T_{ab}=\ph_{;a}\ph_{;b}-\frac{1}{2}g_{ab}\left(\ph_{;c}\ph_{;d}
g^{cd}\right)\;.
\end{equation}
The massless scalar field can be thought of as a stiff 
fluid when the gradient of the scalar 
field $\phi,_\mu$ is timelike. 
Now we are in a position to state and prove the sufficient condition 
for the dispersal of a massless scalar field:
{\em If, during the dynamical evolution of a gravitationally coupled 
massless scalar field, 
a comoving world line along which $\phi,_\mu$ is timelike, approaches the limit 
where $\phi,_\mu$ is null, then dispersal must take place somewhere 
along that worldline.}

In order to see this, first
we note that in the entire region of the spacetime where $\phi,_\mu$ is timelike, 
we can always set up a comoving coordinate system, where 
the massless scalar field behaves 
like a stiff fluid with the equation of state $p=\rho$. So we 
can consider the comoving world lines of this stiff fluid. 
Now let us consider the case that there is such a 
comoving world line which approaches 
the point where $\phi,_\mu$ becomes null. 
Now from the Einstein's equations we know 
that $\mathcal{R}=\phi^{,\mu}\phi,_\mu=-2\rho$. 
This implies that $\mathcal{R}$ and $\rho$ go to 
zero in the limit of approaching any point where $\phi,_\mu$ is 
null. This implies that $\dot{\rho}$ 
must be negative atleast somewhere along the comoving world 
line that we are considering. Since 
$\rho$ and $p$ are positive by construction, from \eqref{conserv}, we conclude 
that $\theta$ would have to be positive whenever $\dot{\rho}$ 
is negative in the comoving patch. 
This implies that $\theta$ has to change sign at least once 
along this comoving world line. The 
changing of sign of the volume expansion from 
negative to positive necessarily denotes 
a dispersal of the scalar field.

We note that this result is independent of any symmetry assumed 
of the spacetime, such as spherical symmetry or self-similarity. Since 
so far we have been considering dispersal of the scalar field 
along a comoving world line, 
the above result does not, by itself, provide any information as 
to whether the scalar 
field will collapse into a blackhole, or disperse away without 
forming one. However, 
in the case when all the comoving world lines approach the 
limit where $\phi,_\mu$ is null, there 
must be a dispersal along all the comoving world lines in that patch, 
and then there would be no blackhole formation. This in fact happens in 
the Roberts solution which we discuss below as an example. 
In the next section, we consider this solution in order to demonstrate 
explicitly, how 
the dispersal of the scalar field accompanies changing of the 
gradient $\phi,_\mu$ from a timelike to a spacelike vector.

Now we state another interesting consequence of the above result,
which is as follows:
{\em If the dynamical evolution of a massless scalar field, starting from a 
spacelike hypersurface on which the gradient of the scalar field is timelike, 
and the volume 
expansion for the comoving congruence is negative to begin with, is such that 
the gradient changes 
from being timelike to null in the future, then there must be a bounce 
or dispersal of the scalar field in the part of the spacetime 
where the gradient is timelike.}

As should be clear, we have here a situation in mind where the scalar field
is gravitationally collapsing initially, when it begins the dynamical evolution.
To see why the above result is true, let us consider the case of the massless 
scalar field, where the dynamical evolution takes place from an initial spacelike 
hypersurface $t=t_i$, and the gradient $\phi,_\mu$ is timelike on that surface. 
This implies that we can construct a comoving congruence evolving from 
the initial hypersurface.

Next, if there is a hypersurface $S$ in the 
future of the initial surface $t=t_i$ where $\phi,_\mu$ is null, this would then 
imply that any comoving congruence defined in the past of $S$ must 
break down at $S$. In other words, the comoving coordinate system describing
the scalar field breaks down at $S$. 
Since $S$ is the boundary of the comoving patch, there are necessarily 
comoving lines which approach arbitrarily close 
to $S$. Then using the earlier result, 
it is seen that there must be dispersal somewhere along any such comoving 
worldline. Therefore 
there must be a dispersal of the scalar field 
in the part of the spacetime where the 
gradient is timelike.

In the argument above, we have implicitly assumed that the density does not go 
to a vanishing value asymptotically on the initial spacelike hypersurface. 
If the density does go to 
zero asymptotically, then it would not be possible 
to argue in the above manner, that 
the density has to decrease from the initial 
value if the gradient of the field becomes 
null in the future. In fact, for the Roberts solution 
discussed below, this turns out 
to be the case.
However, even in that case, the result we mentioned above holds true
as we shall show.
In general also, even when the density asymptotically falls off on a given
spacelike surface, the result above holds true with 
some minor changes. The proof of this would be given elsewhere.

\section{The Roberts solution as an example}

Now we shall consider a particular analytic solution for massless scalar field to demonstrate 
how dispersal of the scalar field arises in this model. 
The metric for this solution to the Einstein's field equations was given by Roberts \cite{Roberts}.
For a massless scalar field, one can construct a comoving coordinate system when the 
gradient of the scalar field is timelike. 
In the Roberts solution, the gradient changes sign, therefore  a single comoving 
coordinate system cannot be used to cover the whole spacetime manifold. 
However, when the continual gravitational collapse starts from an initial
spacelike surface, the gradient of the scalar field is necessarily timelike. 
In that case the equation of state of the massless scalar field 
in the comoving frame is that of a stiff fluid, with $\rho=p$. 

To demonstrate the dispersal of the scalar field, we need only consider the comoving 
patch of the whole manifold where the gradient is timelike. We then show explicitly 
that the continual collapse turns into a dispersal of the field when the 
gradient $\phi,_\mu$ changes from timelike to a spacelike vector in the future
of the initial surface.

Using notations in \cite{Oshiro} the Roberts solution takes 
the following form in double null coordinates
\ba
 ds^2&=& - du dv + R(u,v)^2 d\Omega^2 \non\\
R(u,v)&=& \frac{1}{2} \sqrt{[(1-p^2)v^2-2vu+u^2]}\;,
\ea
and 
\begin{equation}
 \phi=\pm\frac{1}{2} \ln{\frac{(1-p)v-u}{(1+p)v-u}} \;.\label{phi}
\end{equation}
Here the constant $p$ can be chosen to be positive without any loss of generality. 
The units used are such that $8\pi G=c=1$. 
The Ricci scalar for this solution has the form
\begin{equation}
\mathcal{R}= \frac{p^2 u v}{2R(u,v)^4} \label{Ricci}
\end{equation}

\begin{figure}
\begin{center}
\includegraphics[width=5cm]{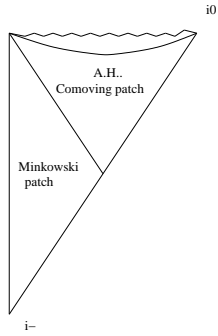}
\caption{\label{fg1}
Penrose Diagram:$p>1$
}
\end{center}
\end{figure}

The spacetime structure depends crucially on the parameter $p$. For $p>1$, 
the singularity becomes timelike in the region $v<0$ and space-like in the region 
$v>0$. For $p=1$ the singularity becomes null. 
For $0<p<1$, the solution has only timelike singularities.

\begin{figure}
\begin{center}
\includegraphics[width=4cm]{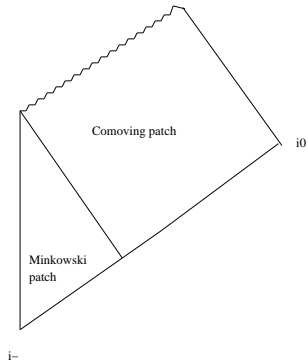}
\caption{\label{fg2}
Penrose Diagram:$p=1$, null singularity
}
\end{center}
\end{figure}


\begin{figure}
\begin{center}
\includegraphics[width=3cm]{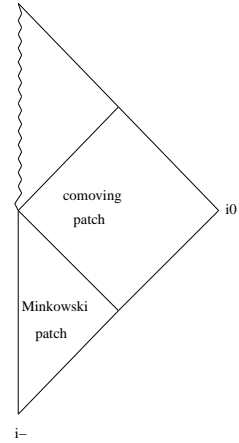}
\caption{\label{fg3}
Penrose Diagram: Solution for $0<p<1$ matched with Minkowski spacetime
}
\end{center}
\end{figure}

Now the singularity in the past can be removed 
by matching the solution with a Minkowski patch, as shown in the figures. 
Then there is an evolution from regular initial data to a future singularity, 
just like a collapse scenario.

\subsection{Transformation into comoving coordinates} 

For massless scalar field, the 
Einstein equations can be written in the form $R_{\mu \nu}= \phi_{,\mu} \phi_{,\nu}$.
>From this it follows that the norm of $\phi_{,\mu}$ is equal to the Ricci scalar. 
Since we would like to construct a comoving coordinate system in the 
part of the manifold where $\phi_{,\mu}$ is timelike,
we have to ensure that the Ricci scalar is negative. From \eqref{Ricci}, it is 
then seen that we must have $uv<0$ in this comoving patch.


Now we consider the solution for which $\phi$ has a negative sign in 
\eqref{phi}.
To make the transformation to the comoving co-ordinates 
$(t,r,\theta,\phi)$, we choose $t(u,v)$ and $r(u,v)$ in such a manner that 
$\phi(u,v)=t$ and $g^c_{tr}=0$, i.e. the comoving metric is diagonal 
and is given in the form 
\be
ds^2=-g^c_{tt}(t,r)dt^2+g^c_{rr}(t,r)dr^2+R^2(t,r)d\Omega^2
\ee
Putting $\phi=t$, in the expression for $\phi$, we have $v=-x(t)u$, 
where $x(t)= [e^{-2t}-1]/[(1+p)(1-e^{-2t})-2p]$.


Using the constraint $g^c_{tr}=0$ and the relation between $v$ and $u$, we can solve explicitly 
for $u(t,r)$ and $v(t,r)$ as  
\begin{equation}
 u(t,r)= \frac{f(r)}{\sqrt{x}}\;\;,\;\; v(t,r)= -f(r) \sqrt{x} \label{uv}
\end{equation}
where $f(r)$ is the arbitrary function of integration.
The metric components in the comoving frame are 
\begin{equation}
 g^c_{tt}=-\frac{1}{4}\frac{f^2(r)}{x^2}\left[\frac{dx}{dt}\right]^2\;\;,\;\; g^c_{rr}= f'^2(r)
\end{equation}
and
\begin{equation}
 R(t,r)^2= \frac{1}{4} f^2(r)[(1-p^2)x+\frac{1}{x}+2] \label{R}
\end{equation}
Also the Ricci scalar can be calculated as
\begin{equation}
 \mathcal{R}= -\frac{8p^2}{f^2(r)} \frac{x^2}{[(1-p^2)x^2+2x+1]^2} \label{Rc}
\end{equation}
The volume expansion for the timelike congruence of comoving shells 
`$r$' is given by $\theta= [2p^2 R_{,t}]/[R]$.

As discussed earlier, the comoving coordinate system breaks down when 
$\phi_{,\mu}$ becomes null. This corresponds to 
$\mathcal{R}=0$. This happens at $x=0$ and as  $x \to \infty$. From the expression for $x$, 
it is seen that $x=0$ implies $t=0$ and $x \to \infty$ implies 
$t\to -\frac{1}{2} \ln{\frac{1-p}{1+p}}$. 
Therefore for $p>1$, the allowed range for $t$ in the comoving patch is $0<t<\infty$ 
and for $0<p<1$, it is $0<t<-\frac{1}{2}\ln{\frac{1-p}{1+p}}$. For all these 
cases, we can take a $t= t_i$ spacelike surface inside the comoving patch and 
consider the dynamic evolution from that surface. From our earlier discussions, we know 
that $\phi,_\mu$ is timelike in the initial spacelike surface. We note here that for all the 
three cases, $u=0$ is another null hypersurface, where $\phi,_\mu$ becomes null. But  
this surface is in the past of $t=t_i$ ( $t_i$ can be chosen in such a way), hence 
not relevant for our purpose.

Without any loss of generality, we can take the initial time $t_i$ to be 
zero. We note that if $\mathcal{R} \to \infty$, then there is a spacetime singularity. 
This occurs at $(1-p^2)x^2+2x+1=0$. 
The solution of this equation is $x=-\frac{1}{1\pm p}$. But only $x=-\frac{1}{1-p}$ 
can be realized for the time to be real. This implies that 
there would be a singularity when $e^{-2t}=0$ or $t \to \infty$. Let us now discuss the 
three cases depending on the value of $p$.

\subsection{The case of $0<p<1$}

This is an interesting case for the Roberts solution in 
which a future timelike singularity develops. In this case, however, $\phi,_\mu$ becomes 
a spacelike vector from a timelike one. The surface $v=0$ is the hypersurface where 
$\phi,_\mu$ is null. Therefore in this case, the continual collapse condition should be 
violated within the comoving patch. In what follows, we shall show 
that this is indeed the case.

In this case, $\theta= \frac{p^2}{4} \frac{f^2(r)}{R^2} 
[(1-p^2)-\frac{1}{x^2}]\frac{dx}{dt}$. From this relation, it is seen that 
$\theta=0$ at $x=\pm \frac{1}{\sqrt{1-p^2}}$. 
Only the positive sign gives a real value of $t$. 
If we denote the instant when $\theta=0$ as $t=t_b$, then we have
$t_b=-\frac{1}{4} \ln{\frac{1-p}{1+p}}$.

Hence, we see that in the co-moving patch, for $t_0<t<t_b$ we have $\theta<0$, 
and hence the massless scalar field is collapsing. However, 
for $t_b<t<-\frac{1}{2}\ln{\frac{1-p}{1+p}}$,
we have $\theta>0$ which imply that the scalar field has bounced 
back from the collapsing 
state at $t=t_b$, and is expanding in this range. 
This shows clearly that the continual collapse condition is violated in the 
comoving patch and also $\phi,_\mu$ becomes null during evolution. This agrees 
with the statement made in the section III.

\subsection{The case of $p>1$}

In this case we see that the singularity, which is spacelike,
lies in the comoving patch. Also all the shells collapse to a spacelike singularity in 
a finite proper time and the final state is necessarily a black hole. The volume expansion for the 
comoving shells remains negative throughout and at the singularity, $\theta(t,r)\to-\infty$. 
Therefore in this case, $\phi,_\mu$ does not become null in the dynamic evolution from the surface 
$t=t_i$ and there is continual collapse starting from $t=t_i$.

\subsection{The case of $p=1$}

In the marginal case of $p=1$, as seen from the Fig 2, 
the singularity is null and $x= \frac{1-e^{-2t}}{2e^{-2t}}$.
Putting $p=1$ in \eqref{Rc}, we get $\mathcal{R}=-\frac{8}{f^2(r)} \frac{x^2}{(2x+1)^2}$.
This shows that $\mathcal{R}$ can diverge only at $x=-\frac{1}{2}$. 
>From the expression for $x$, we find 
however, that this cannot be realised for any $t>0$. 
So the comoving shells $r>0$ never reach the null singularity.
Moreover, since the singularity is null, no timelike or null geodesic can come out of the 
singularity and reach the comoving observer. In this case also, $\phi,_\mu$ does not 
become null in the dynamic evolution from the surface $t=t_i$. In this case, $\theta$ 
is always negative but goes to zero asymptotically as $t \to \infty.$

\section{Discussion}

In the above, we have identified the change of the 
gradient of the scalar field as a sufficient condition for 
the dispersal of the scalar field. We can consider such a 
dispersal of the scalar field either along a single world line, 
or for a local congruence of the same, and this does not 
by itself determine whether the scalar field would 
collapse into a black hole or disperse away as a whole. 
To determine that, we need
to impose additional condition, such as the said 
behaviour should take place along all the world lines. 
We indicated such a scenario in Sec III.

Here we emphasise that this dispersal occurs in 
that region of the spacetime where the gradient of the 
scalar field is timelike and hence no black hole can 
form in that part of the spacetime. So the question 
that whether a black hole can form in that part of the 
spacetime where $\phi,_\mu$
is spacelike is still open. However, there are 
indications that at least for spherically symmetric spacetimes, 
if there is dispersal along all the 
comoving world lines in the part where $\phi,_\mu$ is timelike, 
then no black hole can form 
even in the part of the spacetime where $\phi,_\mu$ is spacelike. 
We are currently investigating 
this issue.

The physical significance of the above is the following:
Consider a massless scalar field that begins collapse from an
initial spacelike surface, and its gradient is then timelike
at all points. The condition we gave then ensures dispersal of 
this initially collapsing field later, as it evolves.
It is possible that the field may refocus again later
even after the dispersal, but that is an issue to be 
considered and examined separately.

We pointed out that the massless scalar field solution 
given by Roberts obeys this result, 
and further, in the case where no dispersal takes place, a black hole forms.
On the other hand, if there is a dispersal of the scalar field along all the
world lines, then a black hole does not form. 
It would be interesting to investigate 
whether the Roberts solution is the only solution which has 
such a behaviour. 
It would be interesting to explore and examine if there are other 
such solutions which behave in a similar manner.  
It is also possible that 
even if the condition for the local dispersal of the scalar field is 
satisfied not along all the world lines, but along a certain fraction of them, 
even in that case the scalar field could disperse away as a whole without 
forming a black hole.

Our study indicates that to gain insight into 
the phenomena of both dispersal of 
scalar fields and collapse, it may be useful to consider 
more closely the gradient change 
of the scalar fields during their dynamical evolution. 
This is important because, if there is indeed such a connection, 
then we can expect to see critical behaviour in such cases. 
Since our result here does not depend on 
any symmetry of the spacetime, this suggests that the 
dispersal phenomena might be a possible feature of massless 
scalar field solutions in general.


\begin{thebibliography}{99}

\bibitem{Christ} Christodoulou D,  Commun. Moth. Phys. {\bf 105} 337 1986; 
Commun.Math. Phys. {\bf 109} 591, 613 1987.


\bibitem{Choptuik} M. W. Choptuik, Phys.Rev.Lett., {\bf 70}, 1, 1993.

\bibitem{Gundlach} C. Gundlach, J. M. Martin-Garcia, 
Living Reviews in Relativity: {\bf 10} 5, 2000.

\bibitem{Roberts} M. D. Roberts, GRG, {\bf 21}, p.907 (1989).

\bibitem{Brady} P. R. Brady, C.Q.G., {\bf 11}, p. 1255 (1994).

\bibitem{Oshiro} Y. Oshiro, K. Nakamura and A. Tomimatsu, Prog. Theor. 
Phys., {\bf 91}, p.1265 (1994).






\bibitem{ellis} G. F. R. Ellis and H. van Elst, Cosmological models 
(Carg\`{e}se lectures 1998), in {\em Theoretical and Observational 
Cosmology\/}, edited by
M. Lachi\`{e}ze-Rey, p. 1 (Kluwer, Dordrecht, 1999);


\end{thebibliography}
\end{document}